\documentclass[11pt]{article}

\usepackage[margin=1in]{geometry}
\usepackage{authblk}
\usepackage{graphicx}
\usepackage{multirow}
\usepackage{amsmath,amssymb,amsfonts}
\usepackage{amsthm}
\usepackage{mathrsfs}
\usepackage[title]{appendix}
\usepackage{xcolor}
\usepackage{textcomp}
\usepackage{booktabs}
\usepackage{comment}
\usepackage{algorithm}
\usepackage{algorithmicx}
\usepackage{algpseudocode}
\usepackage{listings}
\usepackage[numbers,sort&compress]{natbib}
\usepackage[colorlinks=true,linkcolor=blue,citecolor=blue,urlcolor=blue]{hyperref}

\theoremstyle{plain}

\theoremstyle{definition}

\title{Spectral condensation in a finite nonequilibrium atmospheric transition}

\author[1,2]{Dan Zhao$^\dagger$}
\author[1,2]{Yongwen Zhang$^{\dagger,*}$}
\author[3,4]{Teng Liu}
\author[1,2]{Wenqi Liu}
\author[4]{Jingfang Fan}
\author[4,5]{Xiaosong Chen$^*$}

\affil[1]{Faculty of Science, Kunming University of Science and Technology, Kunming 650500, China}
\affil[2]{Yunnan Key Laboratory of Complex Systems and Brain-Inspired Intelligence, Kunming 650500, Yunnan, China}
\affil[3]{Earth System Modelling, School of Engineering and Design, Technical University of Munich, Munich 80333, Germany}
\affil[4]{School of Systems Science and Institute of Nonequilibrium Systems, Beijing Normal University, Beijing 100875, China}
\affil[5]{Institute for Advanced Study in Physics and School of Physics, Zhejiang University, Hangzhou 310058, China}

\date{
\vspace{-1.0em}
\begin{center}
\small
$^\dagger$These authors contributed equally to this work.\\
$^*$Correspondence: Yongwen Zhang (\texttt{zhangyw@kust.edu.cn}); Xiaosong Chen (\texttt{chenxs@zju.edu.cn}).
\end{center}
\vspace{-1.0em}
}

\begin{document}

\maketitle

\begin{abstract}
Order parameters are difficult to define in high-dimensional nonequilibrium systems that lack a Hamiltonian, a thermodynamic limit or an observed control coordinate. Here we show that such transitions can be diagnosed from the spectrum of occupations over data-derived eigen-microstates. We combine Eigen Microstate Theory with a Marchenko--Pastur random-matrix baseline to isolate an emergent sector, whose entropy quantifies competition among statistically significant collective states. As a finite atmospheric realization, we analyse 51 sudden stratospheric warmings in ERA5. The event-aligned ensemble undergoes spectral condensation, decondensation and recondensation: a polar-vortex state dominated by a few eigen-microstates gives way to a high-entropy regime of competing emergent states before selecting a reorganized weak-vortex state. A stochastic wave--mean-flow model, in which upward wave-activity flux provides a reduced control coordinate, reproduces the same entropy maximum, collapse and top-down timing. These results identify emergent-sector entropy as an order-parameter-like, state-based spectral diagnostic for non-Hamiltonian transitions and place polar-vortex breakdown within a broader class of finite nonequilibrium phase reorganizations.
\end{abstract}


\section{Introduction}\label{sec1}

Phase transitions are conventionally identified through a control parameter, an order parameter and, in equilibrium systems, a statistical weight that organizes microscopic configurations into macroscopic phases~\cite{Landau1937,Stanley1971,Goldenfeld1992}. Many observed transitions in natural systems do not provide these objects in this form. They are finite, dissipative and high-dimensional; their trajectories are sampled empirically rather than generated from a known Hamiltonian; and the effective control may be hidden or distributed across several coupled fields. The central challenge is therefore not simply to detect abrupt change, but to define an order-parameter-like state variable directly from observations, without first projecting the system onto a prescribed scalar coordinate~\cite{Hohenberg1977,Cross1993,Hinrichsen2000}.

Existing early-warning diagnostics address one side of this challenge. Variance and lag-1 autocorrelation quantify changes in fluctuation amplitude, memory or recovery rate in a chosen scalar observable, and have been widely used to study tipping and regime shifts~\cite{Scheffer2009,Dakos2008,Lenton2012,Dakos2012}. These measures are valuable when a physically meaningful low-dimensional coordinate is available, but they do not by themselves define the phase state of a high-dimensional system~\cite{Rietkerk2025}. In finite nonequilibrium systems, several collective configurations may coexist and reorganize before a conventional threshold is crossed~\cite{Boers2025}.

Sudden stratospheric warmings (SSWs) provide a finite geophysical realization of this problem. These events, first identified as abrupt wintertime stratospheric warmings and now commonly characterized by rapid weakening or reversal of the polar-night jet, are among the most dramatic examples of polar-vortex breakdown~\cite{Scherhag1952,Butler2015,Baldwin2021}. Their canonical dynamical explanation is nonlinear interaction between upward-propagating planetary waves and the zonal-mean flow~\cite{Matsuno1971,Andrews1987}. Enhanced upward wave activity is a key precursor to extreme stratospheric events~\cite{PolvaniWaugh2004,BirnerAlbers2017}, while preconditioning and internal variability influence whether and how the vortex crosses into a disrupted state~\cite{HoltonMass1976,Plumb1981,ScottPolvani2006,deLaCamara2017,deLaCamara2019}. Individual SSWs differ in vortex geometry, recovery pathway and downward influence~\cite{CP07,Matthewman2009}, yet they share a rapid collective reorganization of the polar-vortex state~\cite{Yasuda2017,Nakamura2020}. This combination of event-to-event diversity and common system-level breakdown makes SSWs a useful atmospheric testbed for asking whether a high-dimensional transition can be diagnosed from observed atmospheric states, rather than only from a prescribed wind-reversal index.

To address this problem, we use Eigen Microstate Theory (EMT)~\cite{Hu2019,Sun2021,Zhang2024pre} as a state-based spectral description. EMT represents a windowed ensemble of observations by singular-vector eigen-microstates and interprets the squared singular values as occupation probabilities. Mathematically, this uses the same spectral information as EOF/PCA. The physical step is to treat the occupation spectrum as an empirical macrostate and to separate statistically significant collective eigen-microstates from the random background using a Marchenko--Pastur random-matrix baseline~\cite{MarchenkoPastur1967,Mehta2004}. The entropy of this emergent sector, $S_{\mathrm{eg}}$, then measures competition among coherent collective states, rather than the total variance or the entropy of all spectral components.

Applying this framework to an ERA5 ensemble of 51 major SSWs, we identify a finite order--disorder--order transition in spectral probability space. Before onset, the vortex state is spectrally condensed onto a small set of dominant eigen-microstates. During the pre-onset transition regime, the emergent sector expands, its occupation hierarchy weakens and $S_{\mathrm{eg}}$ reaches a maximum, indicating competition among multiple coherent vortex configurations. Near wind reversal, the spectrum recondenses into a reorganized weak-vortex state. A stochastic wave--mean-flow model with upward wave-activity flux as an explicit reduced control coordinate recapitulates the same entropy maximum, sharp collapse and top-down timing. Together, these results define an order-parameter-like spectral description of a finite, non-Hamiltonian atmospheric transition.

\section{Spectral transition in the vortex}\label{sec2}

\subsection{Eigen-microstate condensation}

\begin{figure*}[t]
    \centering
    \includegraphics[width=\textwidth]{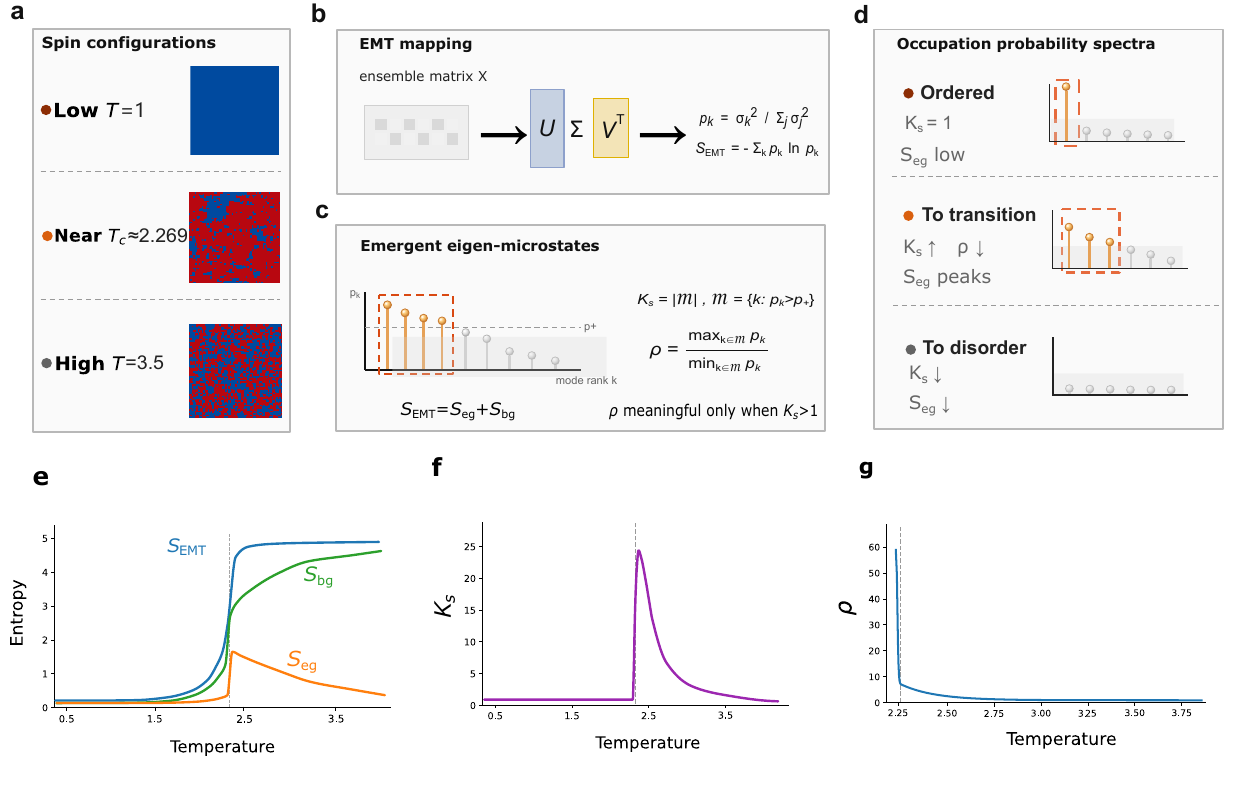}
    \caption{
\textbf{Spectral signatures of phase-transition organization in the EMT--MP framework.}
\textbf{a}, Representative configurations of the two-dimensional Ising model below, near and above the transition regime, showing ordered, heterogeneous and disordered phase organization.
\textbf{b}, EMT maps the ensemble matrix $\mathbf{X}$ to an occupation probability spectrum through SVD, with $p_k=\sigma_k^2/\sum_j\sigma_j^2$ and $S_{\mathrm{EMT}}=-\sum_k p_k\ln p_k$.
\textbf{c}, The Marchenko--Pastur upper edge $p_+$ separates emergent eigen-microstates from the random background, defining $\mathcal{M}$, $K_s=|\mathcal{M}|$ and the hierarchy parameter $\rho$.
\textbf{d}, Schematic occupation spectra illustrating condensation in the ordered regime, competition among multiple eigen-microstates in the transition regime and delocalized/background-dominated occupation in the disordered regime.
\textbf{e}, Entropy decomposition, $S_{\mathrm{EMT}}=S_{\mathrm{eg}}+S_{\mathrm{bg}}$, for the Ising benchmark. The transition-associated peak occurs in the emergent-sector entropy $S_{\mathrm{eg}}$, distinguishing organized eigen-microstate competition from background disorder.
\textbf{f, g}, $K_s$ and $\rho$ quantify the expansion and internal hierarchy of the emergent sector. Together, these diagnostics define a spectral criterion for phase transitions based on the reorganization of eigen-microstate occupations.
}
    \label{fig:figure1}
\end{figure*}

We first use the two-dimensional Ising model as a reference system with known phase organization to establish the physical meaning of the EMT--MP representation. The purpose is not to equate SSW dynamics with an equilibrium spin system, but to calibrate a spectral language for phase transitions that can be applied to both equilibrium ensembles and high-dimensional nonequilibrium trajectories. Representative configurations below, near and above the transition regime show, respectively, an ordered state, spatially heterogeneous domains and a disordered state (Fig.~1a; see Methods).

In the EMT representation, a high-dimensional ensemble is mapped onto an occupation probability spectrum over eigen-microstates (Fig.~1b; see Methods). A strongly organized state corresponds to spectral condensation, in which probability is concentrated onto a small number of dominant eigen-microstates. A weakly organized or disordered state corresponds to a broader distribution over eigen-microstate space. The total eigen-microstate entropy $S_{\mathrm{EMT}}$ therefore measures how delocalized the occupation spectrum is, but it does not by itself distinguish coherent eigen-microstate competition from incoherent background variability.

This distinction is made by combining EMT with a Marchenko--Pastur random-matrix baseline, which separates statistically significant emergent eigen-microstates from the random background (Fig.~1c; see Methods). The entropy can then be decomposed into an emergent-sector contribution $S_{\mathrm{eg}}$ and a background contribution $S_{\mathrm{bg}}$ (Fig.~1e). In the Ising benchmark, $S_{\mathrm{eg}}$ is small in the ordered regime because the spectrum is condensed onto a dominant eigen-microstate. It peaks near the transition regime, where several emergent eigen-microstates coexist and compete for occupation, and decreases again in the disordered regime as the emergent sector loses distinctiveness and occupation leaks into the MP-consistent background. Thus, the phase-transition signal is not the growth of $S_{\mathrm{EMT}}$ alone, but the transient peak of $S_{\mathrm{eg}}$ associated with organized eigen-microstate competition.

The structural diagnostics in Fig.~1f,g quantify the same reorganization. The number of emergent eigen-microstates $K_s$ measures the size of the collective sector, while the hierarchy parameter $\rho$ measures how unevenly occupation is distributed within that sector. Near the transition regime, $K_s$ increases and the occupation hierarchy weakens, indicating that multiple coherent configurations become simultaneously accessible. In the EMT--MP framework, a phase transition is therefore diagnosed as a coordinated redistribution of occupation probabilities across eigen-microstates: condensation in the ordered state, decondensation and competition in the transition regime, and either recondensation or background-dominated delocalization after the transition. This spectral criterion is applied below to diagnose the SSW life cycle.

\subsection{Atmospheric decondensation and recondensation}\label{sec2.2}

\begin{figure*}[t]
    \centering
    \includegraphics[width=\textwidth]{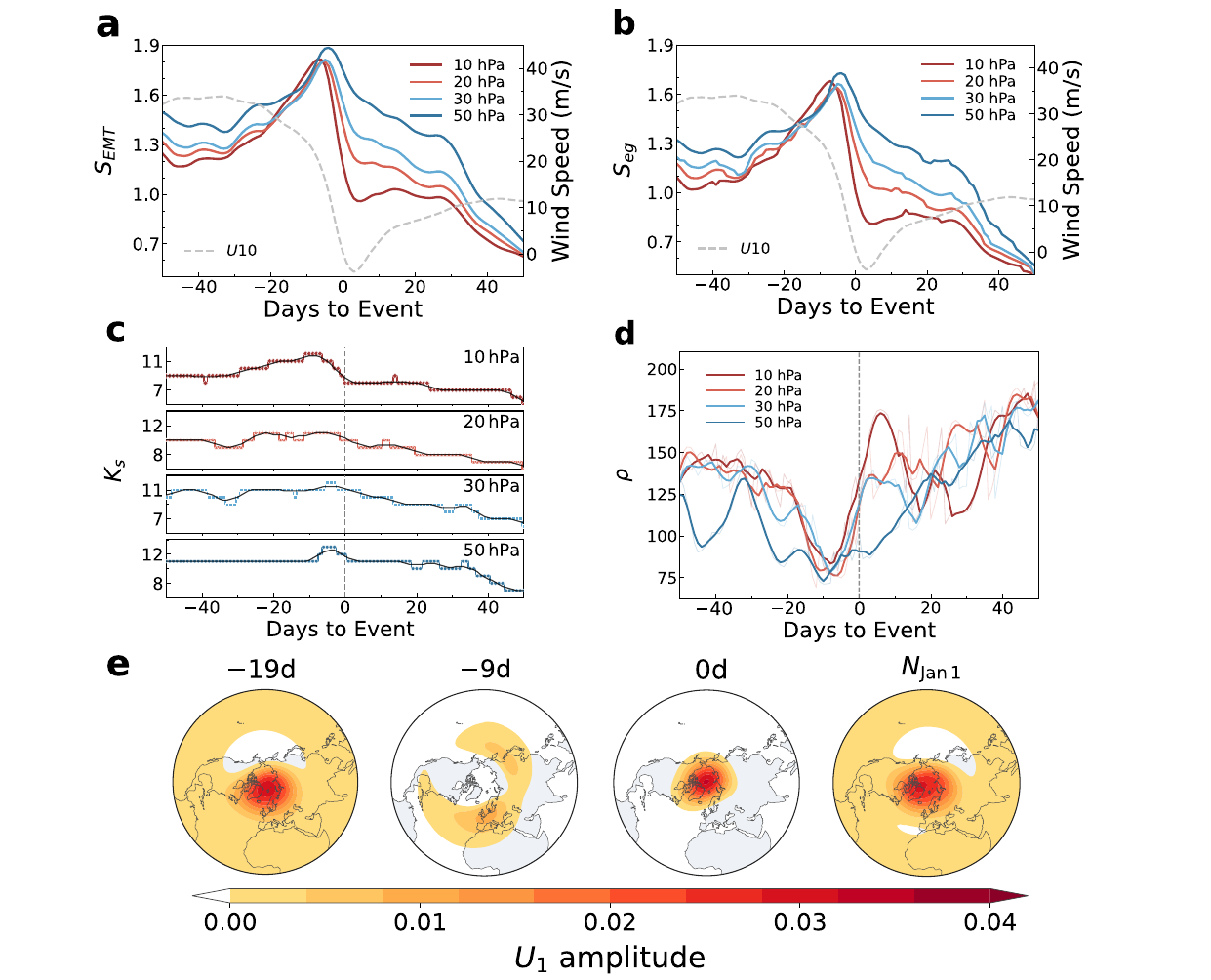}
    \caption{
\textbf{Order--disorder--order phase evolution of SSWs in an event-aligned ensemble.}
\textbf{a}, Composite evolution of the total eigen-microstate entropy $S_{\mathrm{EMT}}$ at 10, 20, 30 and 50\,hPa across 51 major SSWs. Day 0 denotes the CP07 central date, defined by the first reversal of the zonal-mean zonal wind at $60^\circ$N and 10\,hPa; the grey dashed curve shows the corresponding wind index $U_{10}$.
\textbf{b}, Emergent-sector entropy $S_{\mathrm{eg}}$, whose transition-regime maximum precedes the CP07-defined onset and is rooted in statistically significant emergent eigen-microstates.
\textbf{c,d}, Number of emergent eigen-microstates $K_s$ and hierarchy parameter $\rho$, quantifying the expansion and flattening of the emergent sector during the transition regime.
\textbf{e}, Leading spatial eigen-microstate $U_1$ at 10\,hPa as the spatial representation of the dominant phase. The three maps show the pre-transition vortex phase ($-19$\,d), the transition-regime phase near the $S_{\mathrm{eg}}$ maximum ($-9$\,d), and the post-transition SSW phase at the CP07 central date ($0$\,d). $N_{1,\mathrm{Jan}\,1}$ denotes the corresponding non-SSW winter reference phase.
}
    \label{fig:figure2}
\end{figure*}

We next treat the event-aligned ensemble of 51 major SSWs as a finite atmospheric transition. Events are aligned by their Charlton--Polvani (CP07) central dates~\cite{CP07}, denoted by $t_0$, with day 0 defined as the first reversal of the zonal-mean zonal wind at $60^\circ$N and 10\,hPa (see Methods; Supplementary Table S1). Negative lags denote evolution toward onset. For each lag, pressure-level geopotential-height anomalies are pooled over seven daily snapshots before SVD (see Methods).

Figure~2a shows that the total eigen-microstate entropy $S_{\mathrm{EMT}}$ evolves non-monotonically through the SSW life cycle at 10, 20, 30 and 50\,hPa. It increases during the approach to onset, reaches a pronounced maximum several days before the central date, and then decreases sharply near wind reversal. The maximum appears earlier aloft and later below, indicating top-down reorganization. Because this maximum precedes the CP07 wind-reversal date, the spectral transition is not merely a restatement of the onset criterion. In spectral probability space, the vortex first broadens into a high-entropy transition regime and then condenses into a reorganized post-transition SSW phase. Calendar-matched non-SSW winters and entropy-tendency diagnostics confirm that this evolution is event-specific rather than seasonal (Supplementary Figs.~S1--S2).

The entropy decomposition locates the pre-onset maximum in the emergent sector rather than in MP-consistent background variability. The emergent-sector entropy $S_{\mathrm{eg}}$ follows the same non-monotonic evolution as $S_{\mathrm{EMT}}$ at all four pressure levels (Fig.~2b; see Methods). Thus, the peak reflects organized competition among coherent emergent eigen-microstates, not featureless disorder. This is the atmospheric analogue of the Ising benchmark in Fig.~1, where the transition-associated entropy maximum also comes from the emergent sector rather than from total spectral delocalization alone.

The structural origin of this maximum is quantified by $K_s$ and $\rho$ (see Methods). As the system approaches the transition regime, $K_s$ increases as the collective sector expands (Fig.~2c), while $\rho$ decreases as occupation becomes less dominated by a single leading eigen-microstate (Fig.~2d). The transition regime is therefore characterized by decondensation and competition. After the maximum, $K_s$ contracts and the hierarchy sharpens, indicating recondensation onto dominant eigen-microstates associated with the SSW phase.

This phase evolution has a direct spatial representation in the leading spatial eigen-microstate $U_1$ (Fig.~2e; see Methods). Because $U_1$ carries the largest occupation probability, it gives the clearest spatial expression of the selected phase. At $-19$\,d, $U_1$ represents a pre-transition vortex close to the non-SSW winter reference. At $-9$\,d, near the $S_{\mathrm{eg}}$ maximum, it becomes strongly deformed within a broader set of competing emergent eigen-microstates. At day 0, the system has condensed into a reorganized circulation state. Spatial coherence indices and probability distributions further separate these three phases (Supplementary Fig.~S3).

Additional diagnostics support this phase interpretation. Leading spatial eigen-microstates at 20, 30 and 50\,hPa show analogous changes, indicating vertically coherent restructuring of the stratospheric vortex column (Supplementary Fig.~S4). Higher-order eigen-microstates reveal non-axisymmetric displacement-like and wave-2-like structures that populate the transition regime (Supplementary Figs.~S5--S6). These structures explain why the entropy maximum corresponds to competing vortex configurations rather than to an unstructured loss of coherence. The entropy and phase-reorganization signatures are also robust to splitting the SSW sample into independent subsets (Supplementary Figs.~S7--S8).

Thus, the joint evolution of $S_{\mathrm{EMT}}$, $S_{\mathrm{eg}}$, $K_s$, $\rho$ and $U_1$ identifies a finite nonequilibrium transition regime in the SSW life cycle. This phase-state description complements classifications based on vortex morphology, recovery behavior or tropospheric response~\cite{Kodera2016,White2021,Hannachi2025,LuRao2026}: it diagnoses whether the high-dimensional atmospheric ensemble has entered a transition regime.

\subsection{Flux-limited phase selection}\label{sec2.3}

\begin{figure*}[t]
\centering
\includegraphics[width=\textwidth]{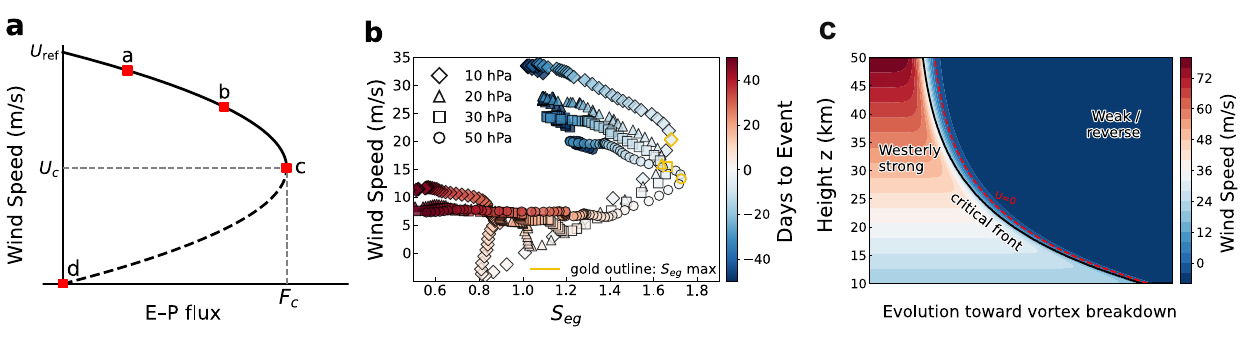}
\caption{
\textbf{Entropy--wind phase portrait and reduced critical-flux interpretation.}
\textbf{a}, Schematic wind--flux relation in the reduced wave--mean-flow threshold framework. The idealized upward wave-activity flux $F$, analogous to an Eliassen--Palm flux, decelerates the vortex from $U_{\mathrm{ref}}$ toward a critical state $(U_c,F_c)$; beyond this point the strong-westerly branch gives way to a weak- or reversed-wind state.
\textbf{b}, ERA5 event-aligned trajectories of emergent-sector entropy $S_{\mathrm{eg}}$ versus zonal-mean zonal wind at $60^\circ$N and 10, 20, 30 and 50\,hPa. Colours indicate days relative to the SSW central date, and gold outlines mark the entropy maximum at each pressure level.
\textbf{c}, Schematic height-dependent critical-flux front. The critical front precedes the zero-wind contour, indicating that loss of the strong-vortex branch can begin before the conventional wind-reversal definition of an SSW is reached.
}
\label{fig:figure3}
\end{figure*}

We next connect the spectral phase reorganization to wave--mean-flow dynamics through a reduced threshold interpretation. Classical SSW theory links vortex weakening to upward-propagating planetary waves and wave-activity or Eliassen--Palm flux convergence~\cite{Matsuno1971,Andrews1987} . Enhanced upward wave activity is a key precursor to weak-vortex and extreme stratospheric events~\cite{PolvaniWaugh2004,BirnerAlbers2017}. The finite-amplitude framework of Nakamura et al.~\cite{Nakamura2020} interprets breakdown as threshold behavior arising from competition between increasing wave activity and decreasing zonal-mean wind: rapid weakening begins once the finite transmission capacity of the mean flow is approached.

We use this idea as a reduced dynamical interpretation of Fig.~2. In the Methods model, wave activity is transported by a mean-flow-dependent group velocity (Methods, Eqs.~\ref{eq:model_A}--\ref{eq:cg}) and weakens the mean flow diagnostically (Methods, Eq.~\ref{eq:mean_flow}), defining an upward flux $F$ (Methods, Eq.~\ref{eq:wave_flux}) as an idealized control coordinate. This construction provides a controlled way to ask whether a flux-limited wave--mean-flow mechanism can generate the observed spectral sequence. ERA5 onset, however, is not assumed to be governed by a unique observed scalar flux; effective control can be distributed across wave forcing, background-flow structure, vertical propagation, preconditioning and internal variability~\cite{HoltonMass1976,Plumb1981,ScottPolvani2006,HitchcockHaynes2016,deLaCamara2017}. Figure~3 therefore combines a reduced theoretical control coordinate with the observed entropy--wind phase portrait, not an inversion of a single atmospheric critical flux.

Within the reduced model, the wind--flux relation has a fold-like structure. Eliminating wave activity from the diagnostic mean-flow relation gives $F\propto u(u_{\mathrm{ref}}-u)$ (Methods, Eq.~\ref{eq:flux_quadratic}), with turning point $u_c=u_{\mathrm{ref}}/2$ and critical flux $F_c$ (Methods, Eq.~\ref{eq:critical_flux}). For weak forcing, the system remains on a strong-westerly branch; as wave activity accumulates, feedback moves it toward the turning point, where this branch loses its ability to transmit additional upward wave activity (Fig.~3a). Beyond this reduced threshold, the system is driven toward weak or reversed winds. The threshold is therefore model-specific, whereas observed breakdown may follow broader wave, flow and preconditioning pathways.

The ERA5 entropy--wind trajectories provide the observational anchor. In the event-aligned phase portrait, $S_{\mathrm{eg}}$ increases as the vortex wind weakens, and its maximum occurs before or near rapid wind collapse at each pressure level (Fig.~3b). The atmosphere therefore enters a weakly selected, high-entropy pre-transition regime before conventional wind-reversal onset, when several coherent eigen-microstates are dynamically accessible. The gold-outlined points mark the entropy maximum at each level, showing a height-dependent approach to transition.

Because the reduced critical flux depends on both the density factor and reference wind (Methods, Eq.~\ref{eq:critical_flux}), the threshold condition is height dependent. Figure~3c summarizes this as a critical-flux front separating the strong-westerly vortex regime from the weak- or reversed-wind regime. In the reduced interpretation, this front precedes the zero-wind contour, so dynamical loss of the strong-vortex branch begins before the CP07 wind-reversal definition is reached. Upper levels approach this flux-limited regime first, followed by progressively lower levels as the transition front migrates downward.

In the EMT--MP description, approaching the reduced flux-limited regime appears as spectral decondensation: the emergent sector expands and the hierarchy flattens, raising $S_{\mathrm{eg}}$ through increasing $K_s$ and decreasing $\rho$ (Methods, Eqs.~\ref{eq:emergent_set}, \ref{eq:rho_hierarchy}, \ref{eq:Seg_Sbg} and \ref{eq:Seg_bound}). Crossing the reduced threshold then reconcentrates the occupation spectrum onto fewer post-onset eigen-microstates. The entropy collapse is therefore phase selection in spectral probability space rather than a simple decay of variability. This mapping separates the reduced flux model, which supplies one explicit dynamical route to rapid vortex breakdown, from $S_{\mathrm{eg}}$, which diagnoses the phase state without requiring the real atmosphere to be controlled by a single scalar coordinate.

\subsection{Stochastic model of recondensation}\label{sec2.4}

\begin{figure*}[t]
    \centering
    \includegraphics[width=\textwidth]{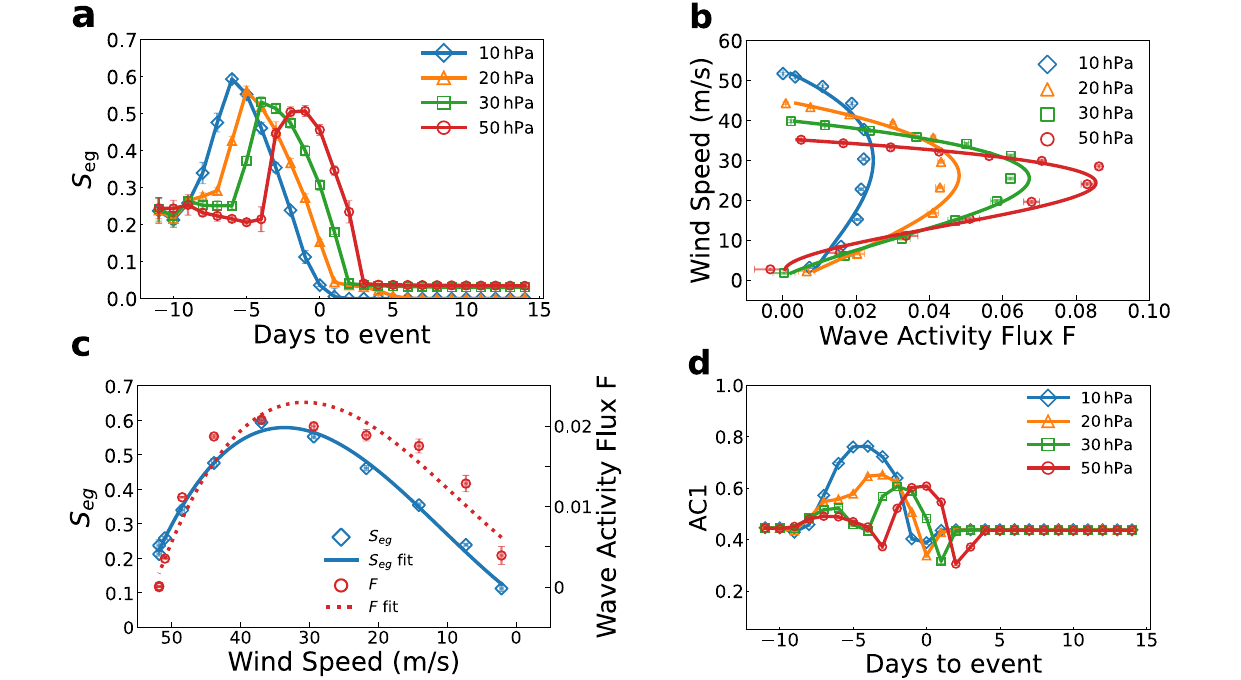}
    \caption{
\textbf{Stochastic wave--mean-flow model reproduces the flux-controlled SSW phase transition.}
\textbf{a}, Emergent-sector entropy $S_{\mathrm{eg}}$ in the stochastic model at four pressure-level analogues. The peak-and-collapse sequence reproduces the order--disorder--order transition and follows a top-down timing consistent with the critical-flux framework.
\textbf{b}, Wind--flux relation showing upward wave-activity flux $F$ as the explicit control coordinate in the reduced stochastic model. The non-monotonic curve reflects the finite transmission capacity of the mean flow.
\textbf{c}, At the 10\,hPa analogue, both $S_{\mathrm{eg}}$ and $F$ peak at intermediate wind speeds, demonstrating that $S_{\mathrm{eg}}$ tracks the flux-controlled transition regime in the reduced model.
\textbf{d}, Lag-1 autocorrelation (AC1) reflects the transition in this low-dimensional setting, but loses diagnostic specificity in ERA5 observations compared with EMT--MP diagnostics (Supplementary Fig.~S11).
}\label{fig:figure4}
\end{figure*}

Having established the critical-flux interpretation, we next test whether this mechanism is sufficient to generate the spectral signatures observed in ERA5. We analyse a stochastic extension of the one-dimensional wave--mean-flow threshold model of Nakamura et al.~\cite{Nakamura2020}, in which the upward wave-activity flux $F$ is explicitly defined (see Methods). The model is not intended to infer a unique scalar control parameter for the real atmosphere. Instead, it provides a controlled dynamical experiment: if the flux-threshold feedback can reproduce the entropy maximum, the subsequent sharp decrease and the top-down timing, then the EMT--MP diagnostics can be interpreted as spectral signatures of a flux-controlled wave--mean-flow transition. We therefore apply the same EMT--MP analysis to the stochastic model ensemble as to the ERA5 event-aligned ensemble (see Methods).

The model reproduces the main temporal structure of the observed SSW transition. At all four pressure-level analogues, $S_{\mathrm{eg}}$ increases before the simulated onset defined from the 10\,hPa analogue (see Methods), reaches a distinct maximum, and then collapses rapidly as the simulated vortex enters a weak- or reversed-wind state (Fig.~4a). The entropy maximum occurs earlier aloft and later below, consistent with the downward progression of the critical-flux front. The corresponding height--time evolution shows rapid wave accumulation aloft, abrupt wind weakening and downward migration of the wind anomaly (Supplementary Figs.~S9--S10). These results show that a flux-threshold feedback is sufficient, in the reduced model, to produce the expansion and collapse of the emergent sector observed in the event-aligned ERA5 ensemble.

The role of \(F\) as the reduced-model control coordinate is made explicit by the simulated wind--flux relation (Methods, Eq.~\ref{eq:flux_quadratic}). As the zonal wind weakens, the upward wave-activity flux first increases, reaches a maximum, and then decreases as the system approaches the weak- or reversed-wind state (Fig.~4b). This non-monotonic relation reflects the finite transmission capacity of the mean flow: once the critical flux is approached, the strong-vortex branch can no longer be continued smoothly and the system is driven toward the post-transition SSW phase. At the 10\,hPa analogue, both $F$ and $S_{\mathrm{eg}}$ peak at intermediate wind speeds, showing that $S_{\mathrm{eg}}$ tracks the transition regime driven by the reduced-model control coordinate $F$ (Fig.~4c).

The model also clarifies the role and limitation of conventional early-warning indicators (see Methods, Eq.~\ref{eq:AC1_def}). In the one-dimensional stochastic model, the lag-1 autocorrelation $\mathrm{AC1}$ increases during the approach to onset and relaxes after the transition (Fig.~4d), consistent with reduced stability near the transition regime. This is expected because the model reduces the transition to a low-dimensional wave--mean-flow instability, for which scalar memory can reflect weakening of the restoring tendency. In high-dimensional ERA5 observations, however, the $\mathrm{AC1}$ signal is much less organized: it is noisier, more sensitive to pressure level and window choice, and does not isolate the spatial or spectral structures involved in the transition (Supplementary Fig.~S11). Thus, $\mathrm{AC1}$ can reflect the transition in the reduced model, but loses diagnostic specificity in the real atmosphere, where the transition is expressed as a collective reorganization of emergent eigen-microstates.

\section{Discussion}

The transition identified here should be interpreted as a finite, spectral reorganization rather than a thermodynamic singularity. SSWs do not provide a controlled thermodynamic limit, and the atmosphere is not sampled from a known Hamiltonian. The useful physical object is therefore the occupation spectrum itself: before onset, probability condenses onto a small set of vortex-like eigen-microstates; during the transition regime, occupation spreads across several emergent collective states; after onset, the spectrum recondenses onto a weak-vortex state. This condensation--decondensation--recondensation sequence is the sense in which the SSW life cycle behaves as a finite nonequilibrium transition.

This framing assigns physical meaning to the occupation spectrum rather than to the singular vectors alone. The singular vectors are mathematically familiar from EOF/PCA, but the relevant phase-state variable is the probability distribution over statistically distinguished emergent eigen-microstates after MP separation. The peak in $S_{\mathrm{eg}}$ therefore identifies competition among collective states, rather than a generic increase in total variance or dimensionality.

The reduced wave--mean-flow model supplies one explicit control coordinate for this spectral reorganization. In the model, upward wave-activity flux drives the system toward a finite transmission capacity of the mean flow, and the same entropy maximum and collapse appear when the threshold is approached and crossed. In ERA5 the effective control is distributed across wave forcing, background vortex structure, vertical coupling and internal variability, so a unique observed scalar coordinate is not required for the diagnostic to be useful. The model instead shows that a known wave--mean-flow threshold mechanism can generate the same spectral order-parameter response seen in the atmospheric ensemble.

These results suggest a different target for predicting atmospheric transitions. Rather than seeking a deterministic single-event metric for the timing of vortex reversal, an ensemble forecast can ask whether the model ensemble has entered a weakly selected phase state in which multiple coherent vortex configurations are simultaneously accessible. In this view, pathway uncertainty is not merely forecast error but part of the transition dynamics: the relevant forecast object is the ensemble-level phase state and its subsequent selection, rather than only the onset date of an individual event. More broadly, the approach suggests a way to define order-parameter-like variables in high-dimensional nonequilibrium systems whose control parameters are hidden, distributed or accessible only through reduced models.

\section{Methods}

\subsection{Reanalysis data and event-aligned ensemble construction}

We analyse daily ERA5 geopotential height and zonal-mean zonal wind during extended boreal winter (November--April) over 1940--2023. The geopotential-height fields are interpolated to a $2.5^\circ \times 2.5^\circ$ Northern Hemisphere grid and analysed independently at 10, 20, 30 and 50\,hPa. At each pressure level, the gridded geopotential-height field is vectorized into a high-dimensional state vector $\mathbf{x}^{(p)}(t)\in\mathbb{R}^{N}$, where $N=144\times37=5328$ is the number of spatial grid points. The reference state $\mathbf{x}_{\mathrm{ref}}^{(p)}$ is defined as the pressure-level-specific climatological mean geopotential-height field over November--April during the 1940/41--2022/23 winters. Anomalies are computed as $\mathbf{y}^{(p)}(t)=\mathbf{x}^{(p)}(t)-\mathbf{x}_{\mathrm{ref}}^{(p)}$. The same climatological reference state is used for both the SSW-centred ensemble and the non-SSW reference analysis.

Major sudden stratospheric warmings (SSWs) are identified following the Charlton--Polvani criterion using the daily zonal-mean zonal wind at $60^\circ$N and 10\,hPa. The central date $t_0$ is defined as the first day on which this wind becomes easterly during the Northern Hemisphere winter season. The additional WMO polar-cap temperature-gradient criterion is not applied. To avoid double counting, no day within 20 days after an identified central date is allowed to define another SSW. Events are classified as final warmings and excluded if the wind becomes easterly but does not return to westerly for at least 10 consecutive days before 30 April. The 51 major SSW central dates used here are listed in Supplementary Table S1. For diagnostics derived from sliding windows, time is referenced to the centre of each window, so that the event-relative lag is
\begin{equation}
\tau = t_c - t_0,
\end{equation}
where $t_c$ is the window-centre day. We analyse 51 major SSWs and construct event-centred composites independently at each pressure level. We distinguish the operational SSW onset from the spectral transition diagnostics. The EMT-defined transition regime is diagnosed independently from the evolution of the occupation spectrum, in particular from the peak of the emergent-sector entropy $S_{\mathrm{eg}}$.

For each event, 7-day sliding windows are constructed with a 1-day step and assigned to their midpoint time. At a given pressure level and lag $\tau$, the 7 daily anomaly fields in the window $[\tau-3,\tau+3]$ are retained for every event and pooled before SVD; no averaging over the event or window is applied. Thus, each SSW event contributes 7 daily snapshots, and the ERA5 SSW ensemble matrix at each lag has $M=51\times7=357$ columns and $N=5328$ rows. For non-SSW reference winters, the same 7-day-window construction is applied using calendar dates rather than SSW central dates. With 41 non-SSW winters, each calendar-centred reference matrix contains $M_{\mathrm{nonSSW}}=41\times7=287$ daily snapshots. The calendar-centred window is advanced by one day to obtain the non-SSW seasonal evolution.

\subsection{EMT framework}

To diagnose nonequilibrium phase transitions in high-dimensional, non-Hamiltonian systems, we employ the eigen-microstate theory (EMT). EMT provides an extended ensemble framework in which system states are represented directly by empirical snapshots in phase space, rather than by a prescribed energy function or a low-dimensional order parameter. All EMT diagnostics below are computed window by window; for notational simplicity, the explicit window and pressure-level indices are suppressed.

Given a high-dimensional state trajectory $\mathbf{x}(t)$, we define the state anomaly as
\begin{equation}
\mathbf{y}(t) = \mathbf{x}(t) - \mathbf{x}_{\mathrm{ref}},
\label{eq:anomaly_def}
\end{equation}
where $\mathbf{x}_{\mathrm{ref}}$ denotes a prescribed reference state. In the ERA5 application, this is the pressure-level-specific climatological reference field defined above.

Sampling the window at times $\{t_m\}_{m=1}^{M}$, we construct the ensemble matrix
\begin{equation}
\mathbf{Y}
=
\big[
\mathbf{y}(t_1),\,\mathbf{y}(t_2),\,\dots,\,\mathbf{y}(t_M)
\big]
\in \mathbb{R}^{N\times M},
\label{eq:ensemble_matrix}
\end{equation}
which contains the microstates visited by the system in the $N$-dimensional phase space. For spectral analysis, $\mathbf{Y}$ is normalized by its Frobenius norm,
\begin{equation}
\widehat{\mathbf{Y}} =
\frac{\mathbf{Y}}{\|\mathbf{Y}\|_F}.
\label{eq:frobenius_norm}
\end{equation}
This normalization fixes the total spectral weight to unity and allows occupation probabilities to be compared across windows.

We apply singular value decomposition (SVD) to the normalized ensemble matrix:
\begin{equation}
\widehat{\mathbf{Y}}
=
\mathbf{U}\boldsymbol{\Sigma}\mathbf{V}^{\top},
\label{eq:svd}
\end{equation}
where
\[
\mathbf{U}
=
[\mathbf{u}_1,\mathbf{u}_2,\dots,\mathbf{u}_r]
\in \mathbb{R}^{N\times r},
\qquad
\mathbf{V}
=
[\mathbf{v}_1,\mathbf{v}_2,\dots,\mathbf{v}_r]
\in \mathbb{R}^{M\times r},
\]
and $r=\mathrm{rank}(\widehat{\mathbf{Y}})$. The columns $\mathbf{u}_k$ and $\mathbf{v}_k$ are the spatial and temporal eigen-microstates, respectively. The diagonal matrix
\begin{equation}
\boldsymbol{\Sigma}
=
\mathrm{diag}(\sigma_1,\dots,\sigma_r)
=
\mathrm{diag}(\sqrt{p_1},\dots,\sqrt{p_r})
\end{equation}
contains the non-zero singular values. Because $\|\widehat{\mathbf{Y}}\|_F^2=1$, the squared singular values define normalized occupation probabilities,
\begin{equation}
p_k=\sigma_k^2,
\qquad
\sum_{k=1}^{r}p_k=1,
\qquad
p_1\ge p_2\ge \cdots \ge p_r>0.
\label{eq:occupation_prob}
\end{equation}

The ensemble matrix can therefore be written as a weighted sum of eigenensembles,
\begin{equation}
\widehat{\mathbf{Y}}
=
\sum_{k=1}^{r}
\sqrt{p_k}\,\mathbf{E}_k,
\label{eq:eigenensemble_sum}
\end{equation}
where
\begin{equation}
\mathbf{E}_k=\mathbf{u}_k\mathbf{v}_k^{\top},
\qquad
(\mathbf{E}_k)_{im}=u_{ik}v_{mk}.
\end{equation}
The weight $p_k$ represents the occupation probability of the $k$-th eigenensemble associated with the corresponding eigen-microstates.

At a mathematical level, this construction is equivalent to EOF/PCA applied to the windowed anomaly snapshots. The conceptual step of EMT is to interpret the modes as eigen-microstates of an effective ensemble and the squared singular values as statistical occupations. The probability vector
\[
\mathbf{p}=(p_1,\dots,p_r)
\]
therefore defines the EMT macrostate in spectral probability space.

To quantify the degree of disorder in this macrostate, we define the eigen-microstate entropy~\cite{liu2025}
\begin{equation}
S_{\mathrm{EMT}}
=
-\sum_{k=1}^{r}p_k\ln p_k.
\label{eq:SEMT}
\end{equation}

The entropy $S_{\mathrm{EMT}}$ quantifies the overall delocalization of the occupation spectrum: ordered states correspond to spectral condensation onto a small number of eigen-microstates, whereas weakly selected or disordered states correspond to broadened occupation spectra. Near a transition regime, competition among multiple eigen-microstates leads to spectral flattening and an entropy increase, indicating collective reorganization of the high-dimensional state space.

\subsection{Spectral identification of phase transitions}

Within EMT, phase transitions are identified through the redistribution of occupation probabilities across eigen-microstates. This definition is not restricted to nonequilibrium systems. In equilibrium systems, such as the Ising model, eigen-microstates are constructed from microscopic configurations sampled from the statistical ensemble. In high-dimensional nonequilibrium systems, the same construction is applied to empirical state-space trajectories. In both cases, the essential signature of a phase transition is a collective reorganization of the occupation spectrum.

An ordered, or condensed, phase is characterized by macroscopic occupation of a subset of eigen-microstates,
\begin{equation}
\sum_{k\in\mathcal{M}}p_k=\mathcal{O}(1),
\label{eq:macroscopic_occupation}
\end{equation}
where $\mathcal{M}$ denotes the set of emergent eigen-microstates with size
\begin{equation}
K_s=|\mathcal{M}|,\qquad K_s\ge1,\qquad K_s\ll r .
\end{equation}
Single-eigen-microstate condensation corresponds to $p_1=\mathcal{O}(1)$ and $K_s=1$, analogous to the ordered phase of the Ising model and conceptually parallel to Bose--Einstein condensation. By contrast, a delocalized macrostate exhibits no macroscopic occupation of any eigen-microstate, with $p_k\approx 1/r$, and is consistent with a noise-dominated or weakly organized background.

A phase transition is therefore represented spectrally as a decondensation, redistribution and possible recondensation of occupation probabilities across eigen-microstates. The transition regime is characterized by weakened occupation hierarchy, activation of additional emergent eigen-microstates and enhanced competition among them. This spectral view provides a common language for both equilibrium phase transitions and high-dimensional nonequilibrium transitions.

To distinguish emergent eigen-microstates from stochastic or finite-sample background variability, we use the Marchenko--Pastur (MP) random-matrix baseline. For an $N\times M$ random matrix with independent entries, the eigenvalue spectrum of the corresponding sample covariance follows the MP law. Because the ensemble matrix is Frobenius-normalized, the squared singular values define occupation probabilities with unit total weight. The corresponding normalized MP bounds are
\begin{equation}
p_{\pm}
=
\frac{(1\pm\sqrt{q})^2}{N},
\qquad
q=\frac{N}{M}.
\label{eq:MP_bounds}
\end{equation}
Here $p_-$ and $p_+$ denote the lower and upper edges of the MP-consistent background. We define the emergent set as
\begin{equation}
\mathcal{M}
=
\{k:p_k>p_+\},
\qquad
K_s=|\mathcal{M}|.
\label{eq:emergent_set}
\end{equation}
The MP criterion is thus used as an operational separation between statistically significant collective eigen-microstates and the random-matrix background; the phase-transition definition itself is based on the reorganization of the occupation spectrum.

The total occupation carried by the emergent sector is $1-\delta$, where
\begin{equation}
\delta
=
1-\sum_{k\in\mathcal{M}}p_k
\label{eq:delta}
\end{equation}
measures leakage into the MP-consistent background. The internal hierarchy of the emergent sector is quantified by
\begin{equation}
\rho
=
\frac{\max_{k\in\mathcal{M}}p_k}
{\min_{k\in\mathcal{M}}p_k}
\ge 1,
\label{eq:rho_hierarchy}
\end{equation}
for $K_s>1$. A large $\rho$ indicates dominance by one emergent eigen-microstate, whereas a smaller $\rho$ indicates stronger competition among emergent eigen-microstates.

The eigen-microstate entropy can be decomposed into emergent-sector and background-sector contributions,
\begin{equation}
S_{\mathrm{EMT}}
=
S_{\mathrm{eg}}+S_{\mathrm{bg}},
\label{eq:entropy_decomposition}
\end{equation}
where
\begin{equation}
S_{\mathrm{eg}}
=
-\sum_{k\in\mathcal{M}}p_k\ln p_k,
\qquad
S_{\mathrm{bg}}
=
-\sum_{k\notin\mathcal{M}}p_k\ln p_k.
\label{eq:Seg_Sbg}
\end{equation}
Introducing normalized probabilities within the two sectors,
\begin{equation}
q_k=\frac{p_k}{1-\delta},
\quad k\in\mathcal{M},
\qquad
t_k=\frac{p_k}{\delta},
\quad k\notin\mathcal{M},
\label{eq:q_t_def}
\end{equation}
we obtain
\begin{equation}
S_{\mathrm{eg}}
=
-(1-\delta)\ln(1-\delta)
+
(1-\delta)H(\mathbf{q}),
\label{eq:Seg_decomp}
\end{equation}
and
\begin{equation}
S_{\mathrm{bg}}
=
-\delta\ln\delta
+
\delta H(\mathbf{t}),
\label{eq:Sbg_decomp}
\end{equation}
where $H(\cdot)$ denotes Shannon entropy.

For $K_s>1$, the entropy of the emergent-sector distribution satisfies
\begin{equation}
H(\mathbf{q})
\ge
\ln K_s-\ln\rho.
\label{eq:Hq_bound}
\end{equation}
Combining Eqs.~(\ref{eq:Seg_decomp}) and (\ref{eq:Hq_bound}), in the low-leakage ordered regime where $\delta$ is negligible, the emergent-sector entropy obeys
\begin{equation}
S_{\mathrm{eg}}
\gtrsim
\ln K_s-\ln\rho.
\label{eq:Seg_bound}
\end{equation}
This bound is used only as an interpretive relation. All entropy diagnostics reported in the main text are computed directly from the EMT occupation spectrum obtained by SVD. The inequality indicates how increasing $K_s$ expands the active emergent sector, while decreasing $\rho$ flattens occupation within that sector. Both effects act to increase $S_{\mathrm{eg}}$ and help interpret the spectral origin of the entropy peak.

As the system crosses the transition, the emergent sector may recondense onto fewer dominant eigen-microstates, reducing $K_s$ and sharpening the hierarchy. Alternatively, the system may evolve toward a more delocalized state, increasing leakage into the MP-consistent background. In either case, $S_{\mathrm{eg}}$ can decrease after the transition, so that the emergent-sector entropy exhibits a peak near the transition regime. The background contribution $S_{\mathrm{bg}}$ is controlled primarily by $\delta$ and reflects the degree to which occupation leaks into the MP-consistent background. Together, $S_{\mathrm{EMT}}$, $S_{\mathrm{eg}}$, $S_{\mathrm{bg}}$, $K_s$, $\rho$ and $\delta$ provide a spectral set of order-parameter diagnostics for phase transitions. 

\subsection{Ising benchmark}

We use the two-dimensional ferromagnetic Ising model as a benchmark system with a known equilibrium transition. The simulations are performed on a square lattice of size $L\times L$, with $L=64$, coupling $J=1$ and periodic boundary conditions in both lattice directions. The critical temperature is $T_c=2.269185314213022$. Spin configurations are generated using a Metropolis Monte Carlo algorithm, with one Monte Carlo sweep consisting of $L^2$ random spin-flip attempts. Temperatures are sampled from $T_{\min}=0.2$ to $T_{\max}=4.0$, with a refined spacing $\Delta T=0.05$ within $T_c\pm0.60$ and a coarser spacing $\Delta T=0.20$ outside this interval.

At each temperature, $M=160$ spin configurations are retained after equilibration. A pilot run is used to estimate the integrated autocorrelation time of the magnetization, and production samples are separated by $\max(200,10\tau_{\mathrm{int}})$ Monte Carlo sweeps. Each $64\times64$ configuration is flattened into a vector of length $N=4096$, producing an ensemble matrix $\mathbf{Y}_{\mathrm{Ising}}(T)\in\mathbb{R}^{4096\times160}$ in the same $N\times M$ convention used above. No additional centering is applied. The matrix is normalized by its Frobenius norm before SVD, and the EMT occupation probabilities are computed as $p_k=\sigma_k^2/\sum_j\sigma_j^2$. The MP upper edge is computed as $p_+=(1+\sqrt{q})^2/N$, with $q=N/M$. Eigen-microstates satisfying $p_k>p_+$ define the emergent sector, from which $K_s$, $S_{\mathrm{EMT}}$, $S_{\mathrm{eg}}$, $\rho$ and $\delta$ are computed directly from the occupation spectrum.

\subsection{Conventional early-warning indicators}

For comparison with conventional one-dimensional early-warning indicators, we compute the lag-1 autocorrelation (AC1) in sliding windows. For ERA5, the scalar diagnostic $z_n$ is the deseasonalized area-mean geopotential-height anomaly averaged over the polar cap $60^\circ$--$90^\circ$N, computed separately at 10, 20, 30 and 50\,hPa. For each major SSW event, a 121-day segment from day $-60$ to day $+60$ relative to the central date is extracted.

Within each window, $z_n$ is linearly detrended and demeaned. The lag-1 autocorrelation is estimated as
\begin{equation}
\mathrm{AC1}_w
=
\frac{
\sum_n \tilde{z}_n\tilde{z}_{n+1}
}{
\left(\sum_n \tilde{z}_n^2\right)^{1/2}
\left(\sum_n \tilde{z}_{n+1}^2\right)^{1/2}
},
\label{eq:AC1_def}
\end{equation}
where $\tilde{z}_n$ denotes the detrended and window-demeaned residual. The corresponding recovery-rate diagnostic is defined as
\begin{equation}
\lambda_w
=
\frac{\ln(\mathrm{AC1}_w)}{\Delta t},
\label{eq:OU_restoring}
\end{equation}
where $\Delta t=1$ day. Values with $\mathrm{AC1}_w\le0$ or $\mathrm{AC1}_w\ge1$ are excluded from the $\lambda_w$ estimate. As marginal stability is approached, $\mathrm{AC1}_w\to1^{-}$ and $\lambda_w\to0^{-}$, reflecting critical slowing down. Sensitivity to window length is assessed using 7-, 10- and 14-day sliding windows (Supplementary Fig.~S11).

These conventional indicators are useful scalar measures of memory and local stability, but they are based on one-dimensional projections. They do not identify which spatial structures participate in the transition, nor do they separate coherent emergent eigen-microstates from background fluctuations. They therefore serve here as a low-dimensional benchmark against which the EMT-based high-dimensional diagnostics are compared.

\subsection{Minimal stochastic wave--mean-flow model}

To isolate the dynamical mechanism underlying the entropy peak-and-collapse sequence, we adopt a one-dimensional stochastic wave--mean-flow model following the threshold framework of Nakamura and co-workers. On a vertical coordinate $z$, the wave-activity amplitude $A(z,t)$ satisfies a conservative transport equation with weak diffusion,
\begin{equation}
\frac{\partial}{\partial t}
\big(\varrho A\big)
=
-\frac{\partial}{\partial z}
\big(\varrho c_g A\big)
+
\kappa \varrho
\frac{\partial^2 A}{\partial z^2},
\label{eq:model_A}
\end{equation}
where
\begin{equation}
\varrho(z)=e^{-z/H}
\end{equation}
is the background density profile. The vertical group velocity is taken to be proportional to the local mean flow,
\begin{equation}
c_g=Cu,
\label{eq:cg}
\end{equation}
and the mean flow is diagnostically coupled to wave activity through
\begin{equation}
u(z,t)
=
u_{\mathrm{ref}}(z,t)-\alpha A(z,t).
\label{eq:mean_flow}
\end{equation}
This closes a positive feedback loop: increasing wave activity decelerates the jet, weaker westerlies reduce upward propagation, and the resulting flux convergence promotes further local wave accumulation.

The upward wave-activity flux is
\begin{equation}
F(z,t)
=
\varrho C u A.
\label{eq:wave_flux}
\end{equation}
Eliminating $A$ using Eq.~(\ref{eq:mean_flow}) gives
\begin{equation}
F(z,t)
=
\frac{\varrho(z)C}{\alpha}
u(z,t)
\big[
u_{\mathrm{ref}}(z,t)-u(z,t)
\big].
\label{eq:flux_quadratic}
\end{equation}
This quadratic relation has a turning point at
\begin{equation}
u_c(z,t)=\frac{u_{\mathrm{ref}}(z,t)}{2},
\end{equation}
with corresponding critical flux
\begin{equation}
F_c(z,t)
=
\frac{\varrho(z)C}{4\alpha}
u_{\mathrm{ref}}^2(z,t).
\label{eq:critical_flux}
\end{equation}
When the imposed upward flux reaches this capacity, the strong-westerly branch can no longer be maintained and the system rapidly transitions toward a weak or reversed-wind state.

The stochastic model is integrated on a vertical domain from 10 to 110\,km with grid spacing $\Delta z=100$\,m and time step $\Delta t=360$\,s. The model parameters are $C=3\times10^{-3}$, $H=7$\,km, $\alpha=0.4$ and $\kappa=10.0$. The reference flow $u_{\mathrm{ref}}(z)$ increases linearly over 10--50\,km and is held constant above. Each integration is run for 50 days, with the first 10 days discarded as spin-up. The lower-boundary wave-activity forcing is prescribed as
\begin{equation}
A(z_0,t)
=
\Delta A
\left[
\tanh\left(\frac{t-t_{\mathrm{on}}}{t_0}\right)+1
\right],
\end{equation}
with $\Delta A=3.13$, $t_0=6$\,h and $t_{\mathrm{on}}=20$ days in raw model time. At the upper boundary, a zero-gradient radiation-type condition is used, $A(z_{\mathrm{top}})=A(z_{\mathrm{top}}-\Delta z)$. The wave-activity equation is advanced explicitly in conservative form for $Ae^{-z/H}$, with centred finite differences for the flux-gradient and diffusion terms.

Stochasticity is introduced as an additive cumulative perturbation to the reference wind, rather than to wave activity or to the lower-boundary forcing. Independent Gaussian innovations are applied at the four diagnostic heights with amplitude $\sigma=0.4$ in the reference experiment. For comparison with ERA5 central dates, model onset is diagnosed a posteriori as the first post-ramp zero crossing of the zonal wind at 31\,km, the model analogue of 10\,hPa. The diagnostic heights 20, 23, 26 and 31\,km are used as rounded log-pressure-height analogues of 50, 30, 20 and 10\,hPa, respectively.

For each noise level, we generate an ensemble of 5000 realizations. EMT is applied to the zonal-wind ensemble separately at each diagnostic height. Unlike ERA5, where the state vector is a gridded geopotential-height anomaly field, the model state vector at height $h$ and time $t$ is the vector of zonal wind across stochastic ensemble members,
\begin{equation}
\mathbf{y}_h(t)
=
\left[
\bar{u}_h^{(1)}(t),\ldots,\bar{u}_h^{(N_{\mathrm{ens}})}(t)
\right]^{\top},
\qquad
N_{\mathrm{ens}}=5000.
\end{equation}
After spin-up, the remaining 40-day segment is analysed using the same 7-day sliding-window construction as in the observations. Within each window, 50 frames per day are sampled uniformly from the 240 model time steps, yielding 350 sampled times per window and 34 sliding-window centres for each diagnostic height. Thus, for each height and window, the model EMT matrix has size $5000\times350$. The four model entropy curves in Fig.~4 are computed independently at the four diagnostic heights.

\bibliographystyle{unsrtnat}
\bibliography{references}

\end{document}